\documentclass[aps,%
               prl,%
               reprint,%
               showpacs,%
               superscriptaddress,%
               longbibliography,%
               nofootinbib%
               ]{revtex4-2}

\usepackage[utf8]{inputenc}
\usepackage[usenames,dvipsnames]{xcolor}
\usepackage{graphicx}
\usepackage{amsfonts}
\usepackage{amssymb}
\usepackage{amsmath}
\usepackage{mathtools}
\usepackage{bm}
\usepackage{dsfont}
\usepackage{braket}
\usepackage{tikz}
\usepackage{txfonts}

\usepackage[pdfusetitle,%
            bookmarks=true,%
            colorlinks,%
            linkcolor=blue,%
            citecolor=blue,%
            urlcolor=blue%
            ]{hyperref}

\newcommand{\eq}[1]{Eq.\thinspace(\ref{#1})}

\newcommand{\fig}[1]{Fig.\thinspace{}\ref{#1}}
\newcommand{\fc}[1]{({#1})}
\newcommand{\figc}[2]{Fig.\thinspace{}\ref{#1}\thinspace{}\fc{#2}}
  
\newcommand{\subfigref}[2]{\hyperref[fig:#1]{\ref*{fig:#1}(#2)}}

\newcommand{\TUM}{\affiliation{Technical University of Munich, TUM School of Natural Sciences, Physics Department, 85748 Garching, Germany}}
\newcommand{\MCQST}{\affiliation{Munich Center for Quantum Science and Technology (MCQST), Schellingstr. 4, 80799 M{\"u}nchen, Germany}}
\newcommand{\Harvard}{\affiliation{Department of Physics, Harvard University, Cambridge, MA 02138, USA}}

\begin{document}

\title{Deconfinement Dynamics of Fractons in Tilted Bose-Hubbard Chains}
\author{Julian Boesl}
\TUM \MCQST
\author{Philip Zechmann}
\TUM \MCQST
\author{Johannes Feldmeier}
\Harvard
\author{Michael Knap}
\TUM \MCQST

\date{\today}

\begin{abstract}
Fractonic constraints can lead to exotic properties of quantum many-body systems.
Here, we investigate the dynamics of fracton excitations on top of the ground states of a one-dimensional, dipole-conserving Bose-Hubbard model. We show that nearby fractons undergo a collective motion mediated by exchanging virtual dipole excitations, which provides a powerful dynamical tool to characterize the underlying ground state phases. We find that in the gapped Mott insulating phase, fractons are confined to each other as motion requires the exchange of massive dipoles. When crossing the phase transition into a gapless Luttinger liquid of dipoles, fractons deconfine. 
Their transient deconfinement dynamics scales diffusively and exhibits strong but subleading contributions described by a quantum Lifshitz model.
We examine prospects for the experimental realization in tilted Bose-Hubbard chains by numerically simulating the adiabatic state preparation and subsequent time evolution, and find clear signatures of the low-energy fracton dynamics. 
\end{abstract}

\maketitle

\textbf{\emph{Introduction.---}}Fractonic systems, in which elementary excitations exhibit restricted mobility, have attracted much interest over recent years~\cite{nandkishore:2019, pretko:2020, gromov:2022, chamon:2005, haah:2011, haah:2013, yoshida:2013, vijay:2015}. 
A prominent example are systems that conserve higher multipole moments of a global $U(1)$ charge~\cite{pretko:2017, pretko:2017a,pretko2017higher,pretko:2018a}. 
Such multipole conservation laws drastically impact nonequilibrium properties, entailing Hilbert space fragmentation~\cite{sala:2020, khemani:2022, sala:2020a}, anomalous diffusion~\cite{gromov:2020, feldmeier:2020, morningstar:2020, zhang2020subdiffusion, moudgalya:2021_mapping} and a slowdown in the spread of information \cite{feldmeier2021crit}. A promising approach to realize such phenomena in experimental setups is the preparation of ultracold atomic gases in tilted optical lattices, whose effective behavior is governed by dipole-conserving Bose- or Fermi-Hubbard models. Experimental realizations of such systems have demonstrated subdiffusive dynamics~\cite{sanchez:2020} as well as Hilbert space fragmentation~\cite{scherg:2021, kohlert:2023} for high-energy initial states. 
At low energies, a duality between fractons and elasticity theory indicates a wealth of possible ground state phases~\cite{pretko:2018, gromov:2019,pretko:2019, kumar:2019, zhai:2019, radzihovsky:2020a, zhai:2021}. Recent theoretical work has explored such low-energy properties in microscopic dipole-conserving lattice models, establishing Mott insulating phases, Luttinger liquids of dipoles, and supersolids~\cite{lake:2022, lake:2023, zechmann:2023, lake:2023a}.
However, preparing and probing such low-energy states in experimental setups remains a significant challenge.

\begin{figure}
    \includegraphics[width=0.98\columnwidth]{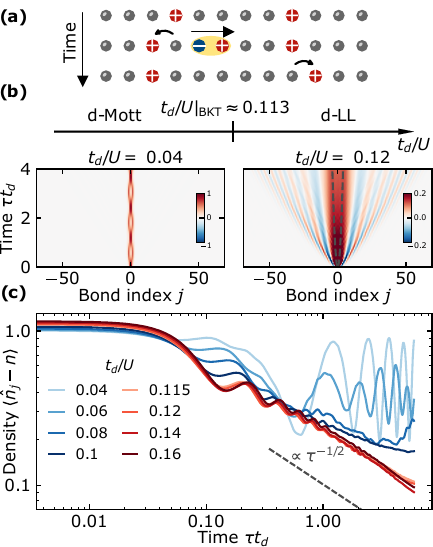}
    \caption{\label{fig:TwoParticles}
        \textbf{Deconfinement of two fractons.} (a) Fractons in dipole moment conserving systems move collaboratively by exchanging dipoles~\cite{pretko:2017b}. (b) Time evolution of the excess density $\langle \hat{n}_j(\tau) - n \rangle$ after adding two particles at adjacent sites on top of the ground state at filling $n = 2$. Deep in the dipole Mott insulator (left), the confined particles follow a breathing motion. In the dipole Luttinger liquid (right), they spread diffusively over accessible time scales in accordance with a semi-classical picture (dashed gray line). (c) Time evolution of the local excess density $\langle \hat{n}_0(\tau) - n \rangle$ on a site where a particle was added. The excess density remains finite for the Mott state but decays diffusively in the Luttinger liquid.
    }
\end{figure}

In this work, we examine dynamical probes of fractonic properties using few-fracton excitations on top of the ground states of a dipole-conserving Bose-Hubbard model. We investigate the collective motion of two initially nearby fractons, mediated by virtual dipole excitations, and study how their mobility depends on the underlying ground-state phase (see also the setup discussed in Ref.~\cite{pretko:2017b}); \fig{fig:TwoParticles}. For the dipole Mott insulator with gapped dipole excitations, fractons remain confined. By contrast, for the gapless dipole Luttinger liquid, kinematic constraints are eased and we analyze the resulting deconfining dynamics both numerically and analytically.
Furthermore, a numerical simulation of adiabatic state preparation demonstrates how the confinement-deconfinement dynamics may be realized with quantum simulators of ultracold atoms in optical lattices. We argue that local dynamical probes are crucial to confirm low-energy dipole-conserving dynamics in lieu of static measurements.

\textbf{\emph{Dipole-conserving Bose-Hubbard model.---}}We consider a one-dimensional model of lattice bosons with a constrained hopping term~\cite{lake:2022, lake:2023,zechmann:2023} of the form
\begin{equation}
    \hat{H} = -t_d \sum_j \left( \hat{b}_j^{\dagger} \hat{b}_{j+1}^2 \hat{b}_{j+2}^{\dagger} + \textrm{h.c.} \right) + \frac{U}{2} \sum_j \hat{n}_j (\hat{n}_j -1),
    \label{eq:DBH}
\end{equation}
where $t_d$ is the strength of the correlated hopping and $U$ a repulsive on-site interaction. This Hamiltonian conserves both the total charge (or particle number) $\hat{N} = \sum_j \hat{n}_j$ and the associated dipole moment (or center of mass) $\hat{P} = \sum_j j \hat{n}_j$. Due to the dipole constraint, single charge excitations created by $\hat{b}_j^\dagger$ act as mobility-restricted fractons, and can move only by emitting or absorbing a mobile dipole excitation $\hat{d}_j^\dagger \equiv \hat{b}_j^\dagger \hat{b}^{}_{j+1}$, see \figc{fig:TwoParticles}{a}.
For a theoretical description of \eq{eq:DBH} at low energies, it is convenient to introduce a local dipole charge $\hat{q}_{d,j}$, defined via $\hat{q}_{d,j} = \sum_{\ell = 0}^j (\hat{n}_\ell - n)$~\cite{moudgalya:2021_mapping, feng:2022_mapping, zechmann:2023}.
Here, $n$ is the average charge density, and we take $n \in \mathbb{N}$ to be integer throughout this work. Crucially, assuming a finite energy gap for single charge excitations, the local dipole charge $q_{d,j}$ remains bounded in the ground state of \eq{eq:DBH}~\cite{zechmann:2023}. 
A standard bosonization procedure gives a counting field $\phi(x)$ and phase field $\theta(x)$ for the fractons, which satisfy $[\partial_x\phi(x), \theta(x^\prime)] = -i\pi \delta(x - x^\prime)$~\cite{giamarchi:2003}. Considering the definition of the dipole density  $\hat{q}_{d,j}$, one can bosonize the dipole degrees of freedom to find the relation between the fracton and the dipole fields $\partial_x \phi_d(x) = \phi(x)$ and $\theta_d(x) = -\partial_x \theta(x)$ leading to equivalent commutation relations $[\partial_x\phi_d(x), \theta_d(x^\prime)] = -i\pi \delta(x - x^\prime)$ (for details see Supplemental Material \cite{supp}).
The effective low-energy description of the system is then given by the sine-Gordon model~\cite{zechmann:2023, lake:2023}  
\begin{equation}
    H_{\text{SG}} = \int \frac{dx}{2\pi} \bigg\{ u_d K_d \left(\partial_x\theta_d\right)^2 + \frac{u_d}{K_d} \left(\partial_x\phi_d\right)^2  + g \cos{\left(\phi_d\right)} \bigg\},
    \label{eq:SGH}
\end{equation}
with Luttinger parameter $K_d$ and Luttinger velocity $u_d$. For $K_d < 2$, realized at small hopping $t_d/U$, the cosine is relevant, pinning the counting field $\phi_d(x)$ and driving the system into a Mott insulator of dipoles with finite mass gap. At a critical hopping strength $t_d/U\bigr|_{\text{BKT}}$, the system undergoes a BKT transition at $K_d = 2$ as the cosine becomes irrelevant. The dipole gap closes and the system enters a Luttinger liquid of dipoles,
\begin{equation}
    H_{\text{LL}} = \frac{u_d}{2\pi} \int dx \left\{ K_d \left(\partial_x\theta_d\right)^2 + \frac{1}{K_d} \left(\partial_x\phi_d\right)^2 \right\}.
    \label{eq:LL}
\end{equation}
Previous numerical studies demonstrated that the lowest integer filling at which a transition into this Luttinger liquid occurs is $n=2$, with $t_d/U\bigr|_{\text{BKT}} \approx 0.113$~\cite{zechmann:2023}. We thus restrict to $n=2$ for the remainder of this work, operating within the phase diagram shown in \figc{fig:TwoParticles}{b}.  

\textbf{\emph{Two-Fracton dynamics.---}}We consider the ground states $\ket{\Omega}$ of the dipole-conserving Bose-Hubbard model \eq{eq:DBH} and add two particles on adjacent sites $\ket{\psi_{2F}} = \hat{b}_0^\dagger \hat{b}_{1}^\dagger \ket{\Omega}$. We note that $\ket{\Omega} = \ket{\Omega \,(t_d/U)}$ depends on the ratio $t_d/U$. Time evolving $\ket{\psi_{2F}}$ under $\hat{H}$, the fractons can hop in opposite directions by the exchange of virtual dipoles acting as `force carriers', reminiscent of mediated interactions in gauge theories~\cite{pretko:2017b,lake:2023a,prakash:2023}; \figc{fig:TwoParticles}{a}. Our goal is to determine the dependence of this dynamical process on the underlying ground state. 

We first discuss the Mott insulating phase. Deep in the strong-coupling limit $t_d/U \ll 1$, the ground state $\ket{\Omega} \approx \ket{222...}$ is close to the homogeneously filled state. The dynamics then takes place in a degenerate subspace spanned by the states $\ket{r} \equiv \hat{b}^\dagger_{-r}\hat{b}^\dagger_{1+r}\ket{\Omega}$, in which the left (right) particle excitation is shifted $r$ sites to the left (right) from its original position. The initial state is given by $\ket{\psi_{2F}} = \ket{r=0}$. 
The degeneracy of this subspace is subsequently lifted by exchanging a single virtual dipole carrying an energy cost $\propto U$. In degenerate perturbation theory, we obtain an effective Hamiltonian
\begin{equation}
    \hat{H}_{2F} = -\sum_{r \geq 0} J_r \ket{r +1}\bra{r} + \textrm{h.c.},
    \label{eq:TwoPartEffec}
\end{equation}
with a position-dependent hopping $J_r \propto t_d^2/U \exp{\left(-r/\xi\right)}$ that decays exponentially over a distance $\xi$ determined by the ratio $t_d/U$ (for details see Supplemental Material~\cite{supp}). The exponential suppression arises as the massive dipole has to travel further to transmit the interaction, dynamically \textit{confining} the two fractons~\cite{pretko:2017b}. 
At very strong repulsion $t_d/U \ll 1$, only the states $\ket{r=0}$ and $\ket{r=1}$ contribute significantly to the dynamics, leading to a periodic breathing motion between these states.
To substantiate this picture of confinement on top of the Mott insulator, even away from $t_d/U \ll 1$, we perform Matrix Product State (MPS) simulations for the model \eq{eq:DBH}. 
We compute the microscopic ground state $\ket{\Omega}$, add two particles on sites $0$ and $1$, and evaluate the time-evolved local excess densities $\braket{\hat{n}_j(\tau) - n} \equiv \braket{\psi_{2F}|e^{i\hat{H}\tau}\,\hat{n}_j \, e^{-i\hat{H}\tau}|\psi_{2F}} - n$. Throughout the Mott insulator, the excess density $\braket{\hat{n}_0(\tau) - n}$ at the initial position of a fracton excitation retains a finite long-time value, in agreement with confinement; \figc{fig:TwoParticles}{c}, blue curves. At very small $t_d/U$, oscillations in $\braket{\hat{n}_0(\tau) - n}$ become apparent. The full spatio-temporal profile of $\braket{\hat{n}_j(\tau) - n}$ shown in \figc{fig:TwoParticles}{b}, left panel, reveals that this is indeed due to the breathing motion of the confined fractons. 

Moving across the phase transition into the dipole Luttinger liquid, the gap of the dipole exchange particles closes, lifting the exponential suppression of the correlated hopping. We thus expect the two fractons to \textit{deconfine} and propagate apart.
In a semi-classical picture, we assume that the rate $dr/d\tau$ at which the distance $r$ between the fractons increases is determined by the time $r/u_d$ it takes a dipole at velocity $u_d$ to travel between them, i.e., $\frac{dr}{d\tau} \propto r^{-1}$.
This leads to a diffusive space-time scaling $r \propto \sqrt{\tau}$. 
We observe dynamics consistent with this semi-classical description on numerically accessible time scales in the diffusive decay of $\langle \hat{n}_0(\tau) -n \rangle \sim 1/\sqrt{\tau}$ throughout the Luttinger liquid; \figc{fig:TwoParticles}{c}, red curves. 
This diffusive transport is reflected in the full profile of the excess density $\braket{\hat{n}_j(\tau)-n}$, which in the center of the system broadens as $\sqrt{\tau}$, see \figc{fig:TwoParticles}{b}, right panel. However, intriguingly, $\braket{\hat{n}_j(\tau)-n}$ further exhibits strong oscillations beyond this feature, spreading behind a ballistically moving light cone and bending in a seemingly diffusive fashion. In order to explain the origin of this feature, we will examine the dynamics of local dipole excitations in the following.

\begin{figure}
    \includegraphics[width=\columnwidth]{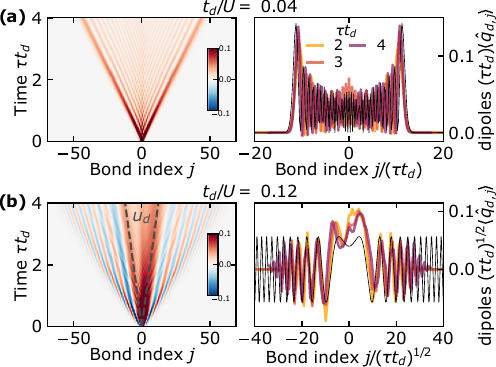}
    \caption{\label{fig:Dipole}
        \textbf{Dynamics of a dipole.} Time evolution of an additional dipole on top of the $n = 2$ ground state. Left column: dipole charge $\langle \hat{q}_{j,d}(\tau) \rangle$. Right Column: Rescaled dipole density cuts at several times. \fc{a} A dipole excitation on top of the Mott insulator expands ballistically and is effectively described by a single free particle (black line, evaluated at $t_d \tau = 4$). \fc{b} A dipole excitation in the Luttinger liquid exhibits pronounced diffusive waves at early times, obeying the Lifshitz scaling relation Eq.~(\ref{eq:LifshitzScaling}) (black line). The late-time dynamics are eventually dominated by the ballistic Luttinger modes (dashed lines).
    }
\end{figure}

\textbf{\emph{Local dipole excitation.---}} In addition to fracton excitations, we can directly study the `force-carrying' dipole excitations by considering the initial state $\ket{\psi_D}=\hat{b}^\dagger_0\hat{b}^{}_1\ket{\Omega}$. 
Deep in the Mott insulator, the effective Hamiltonian governing the dynamics of the dipole excitation corresponds to a single particle nearest-neighbor hopping model (see \cite{supp} for details). The dipole excitation thus spreads ballistically. We confirm this numerically by evaluating the time-evolved local dipole charges $\braket{\hat{q}_{d,j}(\tau)} \equiv \braket{\psi_D|e^{i\hat{H}\tau}\,\hat{q}_{d,j}\,e^{-i\hat{H}\tau}|\psi_D}$, see \figc{fig:Dipole}{a}.

Turning to the dipole Luttinger liquid, the low energy model \eq{eq:LL} predicts two sharp sound modes in the dipole charge $\braket{\hat{q}_{d,j}(\tau)}$, moving right/left with velocity $\pm u_d$ and yielding a dynamical exponent $z=1$. 
Our numerical results indeed indicate the emergence of these sound modes at the latest accessible times, see \figc{fig:Dipole}{b}. The observed dipole density is not inversion symmetric around the origin of the excitation, since the Hamiltonian is not particle-hole symmetric. However, similar to the two-fracton case discussed before, the finite-time dynamics is characterized by additional, strongly oscillating contributions. This suggests the following picture: While \eq{eq:LL} provides the correct \textit{asymptotic} description for late times/low energies, subleading corrections to \eq{eq:LL} are important on accessible, finite times. 

In order to understand these corrections, we recall that \eq{eq:LL} provides the correct low-energy description of the microscopic Hamiltonian \eq{eq:DBH} in the presence of a finite gap for single charge excitations. Previous studies have established a finite charge gap for all $t_d/U$~\cite{lake:2023,zechmann:2023}. However, in practice, this gap can become very small and at finite times the system appears as if charge excitations were gapless. According to the fracton-dipole field relations, the finite charge gap is due to the second term $\sim (\partial_x \phi_d(x))^2 = \phi^2(x)$ in \eq{eq:LL}. Assuming this term is small, we drop it for the purpose of effectively describing early time dynamics. Including the next-to-leading order term $\sim (\partial_x^2 \phi_d(x))^2$ then gives rise to a quantum Lifshitz model~\cite{yuan:2020,lake:2022,radzihovsky:2022},  
\begin{equation}
    H_{\text{Lif}} = \frac{v}{2 \pi} \int dx \left( K (\partial_x \theta_d)^2 + \frac{1}{K} (\partial_x^2 \phi_d)^2 \right),
    \label{eq:Lifshitz}
\end{equation}
which we express in dipole degrees of freedom and where the parameters $v$ and $K$ are named in analogy to the Hamiltonian (\ref{eq:LL}). The energy spectrum follows a quadratic relation $\omega \propto k^2$ and induces a dynamical exponent $z=2$. A recent numerical study of the dipole spectral function in the Luttinger liquid indeed confirmed a quadratic dispersion at higher energies~\cite{zechmann:2023a}. We discuss the relation between the two field theories \eq{eq:LL} and \eq{eq:Lifshitz} in detail in the Supplemental Material~\cite{supp}. 
Using \eq{eq:Lifshitz} as an approximation for early times, we evaluate the time-evolved dipole charge in closed form,
\begin{equation}
        \braket{\hat{q}_{d,j}(\tau)} \propto
    \begin{cases}
      \delta(j), & \tau=0 \\
      \frac{1}{2 \sqrt{v \tau}} \left[ \cos\left( \frac{j^2}{4v \tau} \right) + \sin\left( \frac{j^2}{4v \tau} \right) \right] , & \text{else.}
    \end{cases}
    \label{eq:LifshitzScaling}
\end{equation}
This oscillating function follows a diffusive scaling as expected from the dynamical exponent $z = 2$. 

This expression violates Lieb-Robinson bounds on information spreading, however, causal behavior is restored by a high-momentum cutoff $\Lambda = \mathcal{O}(1/a)$, which in a lattice system is naturally set by the lattice spacing $a$.
The cutoff induces a light cone with finite velocity that approximately corresponds to the group velocity of the quadratic Lifshitz dispersion at the momentum cutoff, $\partial_k \omega(k)|_{\Lambda} = 2v\Lambda$.

Our numerical results for the early-time dipole dynamics agree remarkably well with the scaling relation predicted by the Lifshitz theory; \figc{fig:Dipole}{b}, right column. Also indicated is the Luttinger velocity, extracted from ground state numerics of the Luttinger parameter $K_d$ and the dipole compressibility $\kappa_d$ using the relation $\kappa_d = K_d/u_d \pi$~\cite{zechmann:2023}, which is slow compared to the diffusive Lifshitz oscillations. 
These oscillations are inherited in the two-fracton case discussed previously, and constitute a 
process distinct from the virtual dipole exchange between fractons.

\begin{figure}
    \includegraphics[width=\columnwidth]{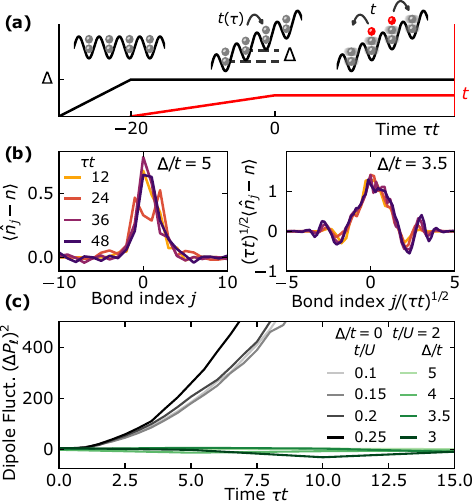}
    \caption{\label{fig:Experiment}
        \textbf{Fractonic dynamics in a tilted lattice.} (a) Sequence for adiabatic preparation of ground states with a two-particle excitation. After a rapid ramp of the tilt $\Delta$, the hopping strength is slowly increased to a finite value $t$. Subsequently, additional particles are introduced for example by optical tweezer potentials. (b) Density profile of the two-particle state for weak (left) and strong (right) final hopping strength. Weak hopping results in the predicted breathing motion. For strong hopping, Lifshitz-like oscillations emerge, with an approximately diffusive scaling. Inversion symmetry is explicitly broken due to the linear potential. (c) Time evolution of dipole moment fluctuations in a segment of size $\ell = 80$ after preparing the excitations on top of an $n = 2$ state. The dipole fluctuations in the tilted lattice do not increase over a significant period of time, suggesting dipole-conserving dynamics (green lines). By contrast, fluctuations on top of a conventional (untilted) $n = 1$ Mott state increase quadratically (gray lines). Fluctuations of adiabatically prepared ground states are subtracted in both cases.
    }
\end{figure}
\textbf{\emph{Experimental realization: Tilted lattices.---}}Having established the dynamics of few-fracton initial states as characteristic signatures of the underlying dipole Mott insulator and Luttinger liquid phases, we now turn to the question how to realize these phases and their dynamical signatures in experiments.  
An accessible platform to implement dipole-conserving dynamics are ultra-cold gases of atoms in an optical lattice with a strong tilt. The Hamiltonian of such a system is given by
\begin{equation}
    \hat{H}  = -t \sum_j \left(\hat{b}^\dagger_j \hat{b}_{j+1} + \textrm{h.c.} \right) + \frac{U}{2} \sum_j \hat{n}_j (\hat{n}_j -1) + \Delta \sum_j j \hat{n}_j,
    \label{eq:HamiltonianTilt}
\end{equation}
where $\Delta$ is the strength of the tilt. In the limit of strong $\Delta \gg t, U$, only correlated processes that conserve the total dipole moment are energetically allowed. A Schrieffer-Wolff transformation yields the dipole-conserving Hamiltonian (\ref{eq:DBH}) with effective correlated hopping $t_{d,\text{eff}} = {t^2 U}/{\Delta^2}$ and a renormalized $U_{\text{eff}} = U (1- {4 t^2}/{\Delta^2})$, alongside a nearest-neighbor interaction of strength ${2t^2 U}/{\Delta^2}$~\cite{scherg:2021, khemani:2022, moudgalya:2022, kohlert:2023}; see Supplemental Material~\cite{supp} for the full derivation. 

The first step is to prepare low-energy states within sectors of fixed dipole moment at integer filling. We propose the following protocol: (i) Initialize the system in a homogeneous state $\ket{222...}$ at integer filling at vanishing hopping $t=0$ and zero tilt $\Delta=0$. (ii) The tilt is then ramped up quickly to a value $\Delta$, leaving the state invariant. This realizes the ground state of the dipole Mott insulator in the limit of vanishing correlated hopping, $t_{d,\mathrm{eff}}=0$. (iii) Next, the depth of the optical lattice is lowered adiabatically, increasing $t$ (and thus $t_{d,\mathrm{eff}}$) until the desired point in the phase diagram is reached; \figc{fig:Experiment}{a}. This results in a state $\ket{\tilde{\Omega}}$ that depends on the final values $t$, $\Delta$, $U$ of hopping, tilt and interactions, as well as the specific adiabatic ramp. 
(iv) Finally, additional particles on top of $\ket{\tilde{\Omega}}$ may be introduced to create the state $\ket{\tilde{\psi}_{2F}} = \hat{b}^\dagger_0 \hat{b}^\dagger_1 \ket{\tilde{\Omega}}$, for example using optical tweezers, see e.g. Refs.~\cite{young2022tweezer,young2023atomic,tao2023high}. Other excitations may be probed as well: Using digital micromirror devices, tunneling between neighboring sites can be induced to access a single-dipole state $\ket{\tilde{\psi}_{D}} = \hat{b}^\dagger_0 \hat{b}_1 \ket{\tilde{\Omega}}$. One can also dope holes in a similar vein; as we can expect hole dynamics over a sufficiently high-filling background to resemble that of particle excitations, this strategy can likewise be used to study single- and few-fracton states.

To demonstrate this protocol, we numerically simulate the adiabatic preparation of $\ket{\tilde{\Omega}}$ and the subsequent dynamics from the two-particle excitation state $\ket{\tilde{\psi}_{2F}}$ using MPS methods. For a given final value $t$ of the single particle hopping, we set $U=0.5t$ and allocate a time $\tau t = 20$ for a linear adiabatic ramp; \figc{fig:Experiment}{a}. We show the dynamics of the excess charge $\braket{\hat{n}_j(\tau)-n}$ from the two-particle state $\ket{\tilde{\psi}_{2F}}$ in \figc{fig:Experiment}{b}. For weak hopping, $\Delta/t=5$, the fractons remain confined with clear signatures of breathing dynamics, distinct from the much faster Bloch oscillations induced by the linear potential. 
In constrast, for larger final hopping, $\Delta/t = 3.5$, we observe dynamical deconfinement of the fractons. The spread of the excess density $\braket{\hat{n}_j(\tau)-n}$ scales approximately diffusively, with strong oscillations reminiscent of the scaling function (\ref{eq:LifshitzScaling}). This suggests that the dynamical properties of the dipole Luttinger liquid -- including strong subleading contributions from a quantum Lifshitz model -- are well captured in this setup.

It remains to verify that the observed diffusive charge dynamics is indeed dipole-conserving. For this purpose, we define the dipole moment $\hat{P}_\ell=\sum_{j=1}^\ell \hat{q}_{d,j-\ell/2}$ in a large linear segment of size $\ell$ around position $j=0$. In experiment, $\hat{P}_\ell$ can be measured from snapshots using quantum gas microscopes~\cite{bakr:2009, sherson:2010}. 
We then consider \textit{fluctuations} of the time-evolved dipole moment $\hat{P}_\ell(\tau)$, which we label as $\bigl( \Delta P^{(2F)}_\ell(\tau) \bigr)^2$ for the initial state $\ket{\tilde{\psi}_{2F}}$, and $\bigl( \Delta P^{(\Omega)}_\ell(\tau) \bigr)^2$ for $\ket{\tilde{\Omega}}$. We note that the latter are non-trivial since $\ket{\tilde{\Omega}}$ is not a true eigenstate. 
The difference $ \bigl(\Delta P_\ell(\tau)\bigr)^2\equiv \bigl(\Delta P^{(2F)}_\ell(\tau)\bigr)^2 - \bigl(\Delta P^{(\Omega)}_\ell(\tau)\bigr)^2$ then quantifies the fluctuation of the dipole moment due to dynamics of charge excitations.
We numerically evaluate the dynamics of $\bigl(\Delta P_\ell(\tau)\bigr)^2$ for a segment of $\ell=80$ and for different tilt-to-hopping ratios $\Delta/t$; \figc{fig:Experiment}{c} (green lines). The fluctuations do not increase, confirming effective dipole conservation on a prethermal time scale. 
By contrast, the dipole fluctuations from two charge excitations on top of a regular $n = 1$ Mott insulator with vanishing tilt $\Delta = 0$ increase rapidly (gray lines). In this case, we predict that the free ballistic movement of the particles leads to $\bigl(\Delta P_\ell(\tau)\bigr)^2 \sim \tau^2$ at late times, consistent with our numerical results.

Finally, one may be tempted to probe the static fluctuations $\bigl(\Delta P^{(\Omega)}_\ell\bigr)^2$ of the state $\ket{\tilde{\Omega}}$ directly: 
For the ground states of the model \eq{eq:DBH} with exact dipole-conservation, these fluctuations scale with $\ell$ as $\bigl(\Delta P^{(\mathrm{dMI})}_\ell\bigr)^2 \sim \mathrm{const.}$ in the dipole Mott insulator, and $\bigl(\Delta P^{(\mathrm{dLL})}_\ell\bigr)^2 \sim \log(\ell)$ in the dipole Luttinger liquid (analogous to particle number fluctuations in a regular Mott state/Luttinger liquid~\cite{song:2010, abanov:2011, rachel:2012}). By contrast, in a regular Mott insulator without dipole-conservation, a finite density of particle-hole fluctuations leads to $\bigl(\Delta P^{(\mathrm{MI})}_\ell\bigr)^2 \sim \ell$, providing a clear distinction to dipole-conserving states.
Crucially however, the tilted model \eq{eq:HamiltonianTilt} enforces dipole-conservation in a rotated basis given by a Schrieffer-Wolff transformation. Since measurements are taken in the standard occupation number basis, this mismatch leads to $\bigl(\Delta P^{(\Omega)}_\ell\bigr)^2 \sim \ell$ despite effective dipole-conservation because of the 'wrong' measurement basis.

\textbf{\emph{Conclusions and Outlook.---}}We have studied the dynamics of local excitations on top of the integer-filling ground states of the dipolar Bose-Hubbard model. Fractons undergo a confinement-deconfinement transition when tuning the initial state from a dipole Mott insulator to a dipole Luttinger liquid. Future work may be dedicated to developing an effective theory of the collective fracton motion and to elucidating its eventual asymptotic late-time behavior. Moreover, it would be interesting to explore the consequences of a modified Mermin-Wagner theorem for our protocols in higher-dimensional dipole-moment conserving systems~\cite{stahl2022spontaneous,lake:2022, stahl:2023}. 

We have furthermore studied the adiabatic preparation and subsequent dynamics of the two-fracton state in a tilted optical lattice setup, identifying dynamical probes as crucial tools to observe fractonic properties at low energies. Our results present clear strategies to realize and probe fractonic low-energy phases. Future studies may explore non-integer commensurate fillings which realize metastable supersolids~\cite{zechmann:2023,lake:2023}. Quasi-two-dimensional gases of polar molecules may offer alternative routes to study fracton deconfinement dynamics, as those systems are effectively described by the elasticity theory of two-dimensional quantum crystals, and in fact supersolid phases have already been demonstrated experimentally~\cite{chomaz:2019}. 

\textbf{\emph{Acknowledgments.---}}We thank Brice Bakkali-Hassani, Immanuel Bloch, Sooshin Kim, and Johannes Zeiher for insightful discussions.
We acknowledge support from the Deutsche Forschungsgemeinschaft (DFG, German Research Foundation) under Germany’s Excellence Strategy--EXC--2111--390814868 and DFG Grants No. KN1254/1-2, KN1254/2-1,  TRR 360 - 492547816, the European Research Council (ERC) under the European Union’s Horizon 2020 research and innovation programme (Grant Agreement No. 851161), as well as the Munich Quantum Valley, which is supported by the Bavarian state government with funds from the Hightech Agenda Bayern Plus.
J.F. acknowledges support by the Harvard Quantum Initiative.
Matrix product state simulations were performed using the TeNPy package~\cite{hauschild:2018}.

\textbf{\emph{Data and Code availability.---}}Numerical data and simulation codes are available on Zenodo upon reasonable request~\cite{zenodo}.

\bibliography{references}
\newpage
\leavevmode \newpage

\setcounter{equation}{0}
\setcounter{page}{1}
\setcounter{figure}{0}
\renewcommand{\thepage}{S\arabic{page}}  
\renewcommand{\thefigure}{S\arabic{figure}}
\renewcommand{\theequation}{S\arabic{equation}}
\onecolumngrid
\begin{center}
\textbf{Supplemental Material:}\\
\textbf{Deconfinement Dynamics of Fractons in Tilted Bose-Hubbard Chains}\\ \vspace{10pt}
Julian Boesl$^{1,2}$, Philip Zechmann$^{1,2}$, Johannes Feldmeier$^{3}$, and Michael Knap$^{1,2}$ \\ \vspace{6pt}

$^1$\textit{\small{Technical University of Munich, TUM School of Natural Sciences, Physics Department, 85748 Garching, Germany}} \\
$^2$\textit{\small{Munich Center for Quantum Science and Technology (MCQST), Schellingstr. 4, 80799 M{\"u}nchen, Germany}} \\
$^3$\textit{\small{Department of Physics, Harvard University, Cambridge, MA 02138, USA}}
\vspace{10pt}
\end{center}
\maketitle
\twocolumngrid

\subsection{A. Derivation of effective Hamiltonians for the dipole Mott state}
In this section, we  derive  effective Hamiltonians for the dynamics of excitations deep in the dipole Mott state of the dipole-conserving Bose-Hubbard model (\ref{eq:DBH}), which we restate for completeness here: 

\begin{equation}
    \begin{aligned}
    \hat{H} &=  \hat{H}_{\text{kin}} +  \hat{H}_U \\  &= -t_d \sum_j \left( \hat{b}_j^{\dagger} \hat{b}_{j+1}^2 \hat{b}_{j+2}^{\dagger} + \textrm{h.c.} \right)  +  \frac{U}{2} \sum_j \hat{n}_j (\hat{n}_j -1).
    \end{aligned}
    \label{eq:DBHAppendix}
\end{equation}
We now consider the strong-interaction limit $t_d/U \ll 1$ at filling $n = 2$. In this case, the ground state $\ket{\Omega}$ is very close to the homogeneously filled product state, $\ket{\Omega} \approx \ket{222...}$. 

The first effective Hamiltonian we consider governs the time evolution of a single dipole excitation on top of the ground state. In this case, the space of possible states is spanned by a set of states $\ket{j_d} = \hat{b}^\dagger_j \hat{b}_{j+1}\ket{\Omega}$, where a single dipole sits on the $j$-th bond in the system. Other configurations are separated from this space due to dipole moment conservation and the fact that further excitations are prohibited by the strong on-site repulsion. To first order, the states $\ket{j_d}$ are connected by the hopping term $\hat{H}_{\text{kin}}$, realizing a nearest-neighbor hopping for the dipole. We can directly evaluate all transition elements as
\begin{equation}
    \bra{j_d}\hat{H}_{\text{kin}}\ket{i_d} = \bra{
\begin{tikzpicture}
\filldraw [black] (0,0) circle (1.5pt);
\filldraw [black] (0.2,0) circle (1.5pt);
\filldraw [black] (0.2,0.15) circle (1.5pt);
\filldraw [black] (0.2,0.3) circle (1.5pt);
\filldraw [black] (0.4,0) circle (1.5pt);
\filldraw [black] (0.4,0.15) circle (1.5pt);
\end{tikzpicture}
}\hat{H}_{\text{kin}}\ket{
\begin{tikzpicture}
\filldraw [black] (0,0) circle (1.5pt);
\filldraw [black] (0,0.15) circle (1.5pt);
\filldraw [black] (0.2,0) circle (1.5pt);
\filldraw [black] (0.4,0) circle (1.5pt);
\filldraw [black] (0.4,0.15) circle (1.5pt);
\filldraw [black] (0.4,0.3) circle (1.5pt);
\end{tikzpicture}
} \delta_{j,i \pm 1} = -6t_d \delta_{j,i \pm 1}.
    \label{eq:DipoleTransitionElement}
\end{equation}
The effective Hamiltonian can therefore be diagonalized in momentum modes, as
\begin{equation}
    \begin{aligned}
    \hat{H}_{1\text{Dip}} = & -6 t_d \sum_j  \ket{(j+1)_d}\bra{j_d} + \textrm{h.c.} \\ = & -12 t_d \sum_k \cos{(k)} \ket{k_d}\bra{k_d},
    \end{aligned}
    \label{eq:HeffDipole}
\end{equation}
where $\ket{k_d} = \frac{1}{\sqrt{L}} \sum_j e^{-ikj} \ket{j_d}$. From this, the time evolution of a local dipole in the middle of the chain, $\ket{\psi_D(\tau = 0)} = \ket{0_d}$, can be derived exactly, which in the continuum limit yields:
\begin{widetext}
\begin{equation}
    \begin{aligned}
    \ket{\psi_D(\tau)} = e^{-i\hat{H}_{1\text{Dip}} \tau}\ket{0_d} & = e^{-i\hat{H}_{1\text{Dip}} \tau} \frac{1}{\sqrt{L}} \sum_k \ket{k_d} = \frac{1}{\sqrt{L}} \sum_k e^{i12 t_d \tau \cos{(k)}} \ket{k_d} = \frac{1}{L} \sum_{k,j} e^{-i(kj - 12t_d \tau \cos{(k)})} \ket{j_d} \\ & \rightarrow \sum_j \int \frac{dk}{2\pi} e^{-i(kj - 12t_d \tau \cos{(k)})} \ket{j_d} = \sum_j J_j(12t_d\tau) \ket{j_d},
    \end{aligned}
    \label{eq:TEMottAppendix}
\end{equation}
\end{widetext}
where $J_j$ is the Bessel function with integer index $j$.

The strong-coupling limit also allows for a perturbative treatment of a state where we put two particles on adjacent sites, $\ket{\psi_{2F}} = \hat{b}_0^\dagger \hat{b}_1^\dagger \ket{\Omega}$. The strong repulsive interactions severely restrict the low-energy subspace in which this state lies, which is furthermore split into distinct sectors with different dipole moment. No further excitations from the Mott state are possible, while conservation of the center of mass implies that the state in question can only be connected to states that can be accessed by moving one particle by $r$ sites to the right, moving the second one by $r$ sites to the left to conserve the dipole moment. We label these states using this integer $r$; $\ket{r}$ is therefore the state in which the distance in bonds of the two particles is $2r +1$, with $\ket{\psi_{2F}} = \ket{r = 0}$ being the initial state.

These states are connected to each other by the exchange of virtual dipoles: A particle may hop into one
direction by emitting a dipole; this dipole may then travel to the other particle and be absorbed, returning to the original low-energy subspace. In this section, we will only consider the lowest-order transition for each state; this is the process where a state $\ket{r}$ is connected to $\ket{r \pm 1}$ by exchange of a single virtual dipole, which hops directly to the second particle without further ado. Thus, higher-order processes related to the emission of multiple dipoles are neglected. 

For the transition from $\ket{r = 0}$ to $\ket{r = 1}$, we only have one intermediary state. In this case, we can apply non-degenerate perturbation theory to obtain the transition element $\bra{r = 1}\hat{H}_{2F}\ket{r = 0}$:

\begin{equation}
\bra{
\begin{tikzpicture}
\filldraw [black] (0,0) circle (1.5pt);
\filldraw [black] (0,0.15) circle (1.5pt);
\filldraw [black] (0,0.3) circle (1.5pt);
\filldraw [black] (0.2,0) circle (1.5pt);
\filldraw [black] (0.4,0) circle (1.5pt);
\filldraw [black] (0.4,0.15) circle (1.5pt);
\filldraw [black] (0.4,0.3) circle (1.5pt);
\filldraw [black] (0.4,0.45) circle (1.5pt);
\filldraw [black] (0.6,0) circle (1.5pt);
\filldraw [black] (0.6,0.15) circle (1.5pt);
\end{tikzpicture}
}\hat{H}_{\text{kin}}\ket{
\begin{tikzpicture}
\filldraw [black] (0,0) circle (1.5pt);
\filldraw [black] (0,0.15) circle (1.5pt);
\filldraw [black] (0.2,0) circle (1.5pt);
\filldraw [black] (0.2,0.15) circle (1.5pt);
\filldraw [black] (0.2,0.3) circle (1.5pt);
\filldraw [black] (0.4,0) circle (1.5pt);
\filldraw [black] (0.4,0.15) circle (1.5pt);
\filldraw [black] (0.4,0.3) circle (1.5pt);
\filldraw [black] (0.6,0) circle (1.5pt);
\filldraw [black] (0.6,0.15) circle (1.5pt);
\end{tikzpicture}
} =
\bra{
\begin{tikzpicture}
\filldraw [black] (0,0) circle (1.5pt);
\filldraw [black] (0,0.15) circle (1.5pt);
\filldraw [black] (0,0.3) circle (1.5pt);
\filldraw [black] (0.2,0) circle (1.5pt);
\filldraw [black] (0.2,0.15) circle (1.5pt);
\filldraw [black] (0.4,0) circle (1.5pt);
\filldraw [black] (0.4,0.15) circle (1.5pt);
\filldraw [black] (0.6,0) circle (1.5pt);
\filldraw [black] (0.6,0.15) circle (1.5pt);
\filldraw [black] (0.6,0.3) circle (1.5pt);
\end{tikzpicture}
}\hat{H}_{\text{kin}}\ket{
\begin{tikzpicture}
\filldraw [black] (0,0) circle (1.5pt);
\filldraw [black] (0,0.15) circle (1.5pt);
\filldraw [black] (0,0.3) circle (1.5pt);
\filldraw [black] (0.2,0) circle (1.5pt);
\filldraw [black] (0.4,0) circle (1.5pt);
\filldraw [black] (0.4,0.15) circle (1.5pt);
\filldraw [black] (0.4,0.3) circle (1.5pt);
\filldraw [black] (0.4,0.45) circle (1.5pt);
\filldraw [black] (0.6,0) circle (1.5pt);
\filldraw [black] (0.6,0.15) circle (1.5pt);
\end{tikzpicture}
} = -6 \sqrt{2}t_d
\label{eq:0to1}
\end{equation}
The energy difference between the excited state and the low-energy subspace is $\Delta E = 2U$. With an additional factor of $2$ as we have two possible intermediary states, the total transition element to second order is $\bra{r = 1}\hat{H}_{2F}\ket{r = 0} = -72 \frac{t_d^2}{U}$.

For larger distances, there are more intermediary states as the virtual dipole has to pass from one fracton to the other. If we consider only one direction in the transition from $\ket{r}$ to $\ket{r+1}$, $2r+1$ states will be involved in the effective interaction, which corresponds to all possible positions of the dipole between the fractons. Labeling these as $\ket{r,j}$, where $j = 1,\ldots,2r+1$ denotes the distance of the dipole from the emitting fracton, it is important to note that all $\ket{r,j}$ for $j \neq 2r +1$ are degenerate under $\hat{H}_U$, with an energy difference to the low-energy subspace of $\Delta E = U$. Therefore, one has to apply degenerate perturbation theory. The perturbation $\hat{H}_{\text{kin}}$ lifts the degeneracy as it couples adjacent states. The problem is equivalent to a free particle on a chain of length $2r$ with open boundary conditions, leading to new eigenstates of the form $\ket{r,k} = \mathcal{N}_k^{-1} \sum_j^{2r} \sin{(kj)} \ket{j}$ with $k = \frac{\pi}{2r+1} i$ for $i = 1,\ldots,2r$ and energy difference $\Delta E = U -12 t_d \cos(k)$, where $\mathcal{N}_k = \sqrt{\sum_j^{2r} \sin^2{(kj)}}$ is a normalization constant. The rightmost state $\ket{r, 2r+1}$ has a higher energy difference $\Delta E = 2U$ and thus can be treated separately. Again taking into account both directions, the full expression for the transition element is

\begin{equation}
    \begin{split}
    &\bra{r+1}\hat{H}_{2F}\ket{r} = \\ & = 2 \sum_k \Biggl\{ \frac{\bra{r, j= 2r+1}\hat{H}_{\text{kin}}\ket{r, k} \bra{r, k}\hat{H}_{\text{kin}}\ket{r}}{2U(U -12t_d\cos(k))} \times \\
    & \qquad \qquad \qquad \qquad \qquad \times \bra{r+1}\hat{H}_{\text{kin}}\ket{r, j= 2r+1} \Biggr\}.
    \end{split}
    \label{eq:DPT}
\end{equation}
All transition elements can be evaluated directly and given in a concise form. The respective expressions are

\begin{widetext}
    \begin{equation}
        \begin{aligned}
            & \bra{r, k}\hat{H}_{\text{kin}}\ket{r} = \mathcal{N}_k^{-1} \sin{(k)} \bra{r, j= 1}\hat{H}_{\text{kin}}\ket{r} = \mathcal{N}_k^{-1} \sin{(k)} \bra{
\begin{tikzpicture}
\filldraw [black] (0,0) circle (1.5pt);
\filldraw [black] (0,0.15) circle (1.5pt);
\filldraw [black] (0,0.3) circle (1.5pt);
\filldraw [black] (0.2,0) circle (1.5pt);
\filldraw [black] (0.4,0) circle (1.5pt);
\filldraw [black] (0.4,0.15) circle (1.5pt);
\filldraw [black] (0.4,0.3) circle (1.5pt);
\end{tikzpicture}
}\hat{H}_{\text{kin}}\ket{
\begin{tikzpicture}
\filldraw [black] (0,0) circle (1.5pt);
\filldraw [black] (0,0.15) circle (1.5pt);
\filldraw [black] (0.2,0) circle (1.5pt);
\filldraw [black] (0.2,0.15) circle (1.5pt);
\filldraw [black] (0.2,0.3) circle (1.5pt);
\filldraw [black] (0.4,0) circle (1.5pt);
\filldraw [black] (0.4,0.15) circle (1.5pt);
\end{tikzpicture}
} = - \mathcal{N}_k^{-1} 3\sqrt{6}t_d \sin{(k)} \\ & \bra{r, j= 2r+1}\hat{H}_{\text{kin}}\ket{r, k} = \mathcal{N}_k^{-1} \sin{(2rk)} \bra{r, j= 2r+1}\hat{H}_{\text{kin}}\ket{r, j= 2r} = \mathcal{N}_k^{-1} \sin{(2rk)} \bra{
\begin{tikzpicture}
\filldraw [black] (0,0) circle (1.5pt);
\filldraw [black] (0,0.15) circle (1.5pt);
\filldraw [black] (0.2,0) circle (1.5pt);
\filldraw [black] (0.4,0) circle (1.5pt);
\filldraw [black] (0.4,0.15) circle (1.5pt);
\filldraw [black] (0.4,0.3) circle (1.5pt);
\filldraw [black] (0.4,0.45) circle (1.5pt);
\end{tikzpicture}
}\hat{H}_{\text{kin}}\ket{
\begin{tikzpicture}
\filldraw [black] (0,0) circle (1.5pt);
\filldraw [black] (0.2,0) circle (1.5pt);
\filldraw [black] (0.2,0.15) circle (1.5pt);
\filldraw [black] (0.2,0.3) circle (1.5pt);
\filldraw [black] (0.4,0) circle (1.5pt);
\filldraw [black] (0.4,0.15) circle (1.5pt);
\filldraw [black] (0.4,0.3) circle (1.5pt);
\end{tikzpicture}
} = - \mathcal{N}_k^{-1} 4\sqrt{3}t_d \sin{(2rk)} \\ & \bra{r+1}\hat{H}_{\text{kin}}\ket{r, j = 2r+1} = \bra{
\begin{tikzpicture}
\filldraw [black] (0,0) circle (1.5pt);
\filldraw [black] (0,0.15) circle (1.5pt);
\filldraw [black] (0.2,0) circle (1.5pt);
\filldraw [black] (0.2,0.15) circle (1.5pt);
\filldraw [black] (0.4,0) circle (1.5pt);
\filldraw [black] (0.4,0.15) circle (1.5pt);
\filldraw [black] (0.4,0.3) circle (1.5pt);
\end{tikzpicture}
}\hat{H}_{\text{kin}}\ket{
\begin{tikzpicture}
\filldraw [black] (0,0) circle (1.5pt);
\filldraw [black] (0.2,0) circle (1.5pt);
\filldraw [black] (0.2,0.15) circle (1.5pt);
\filldraw [black] (0.2,0.3) circle (1.5pt);
\filldraw [black] (0.2,0.45) circle (1.5pt);
\filldraw [black] (0.4,0) circle (1.5pt);
\filldraw [black] (0.4,0.15) circle (1.5pt);
\end{tikzpicture}
} = -6\sqrt{2}t_d.
        \end{aligned}
    \end{equation}
\end{widetext}
The explicit form of the transition element for two states with $r \neq 0$ is 
\begin{equation}
\begin{aligned}
    & \bra{r+1}\hat{H}_{2F}\ket{r} = \\ & -36 \frac{t_d^2}{U} \sum_i^{2r} \frac{\sin{(\frac{2r\pi}{2r+1} i)} \sin{(\frac{\pi}{2r+1} i)}}{\left(\sum_j^{2r} \sin^2{(\frac{\pi}{2r+1} ij)}\right)\left(\frac{U}{12t_d} - \cos{(\frac{\pi}{2r+1} i)}\right)}
\end{aligned}
\end{equation}
Therefore, the effective Hamiltonian of the two-fracton states in this dipole sector can be written as
\begin{equation}
    \hat{H}_{2F} = -\sum_{r \geq 0} J_r \ket{r +1}\bra{r} + \textrm{h.c.},
    \label{eq:H2FAppendix}
\end{equation}
where we have introduced the position-dependent hopping strengths $J_r = - \bra{r+1}\hat{H}_{2F}\ket{r}$. These can be evaluated numerically, which yields a strong exponential suppression of the hopping strength at high distances, of the form $J_r \propto t_d^2/U \exp{\left(-r/\xi\right)}$ for some correlation length $\xi$ which depends on the ratio $t_d/U$, as expected from the physical picture of a massive dipole as a virtual interaction carrier. The effective Hamiltonian (\ref{eq:H2FAppendix}) is reminiscent of a single particle on a semi-infinite chain, the mass of which increases exponentially with distance. In particular, at very strong interaction $t_d/U \ll 1$ the states $\ket{r = 0}$ and $\ket{r = 1}$ are strongly energetically separated from the higher states due to the decay of the coupling, effectively spanning a two-state low-energy subspace. This structure leads to the breathing motion of the initial state considered in the main text, which is confirmed by numerical studies.

\subsection{B. Low-energy charge and dipole field theories}
The defining relation $\hat{q}_{d,j} = \sum_{l = 0}^j (\hat{n}_l - n)$ between the charge density $\hat{n}_j$ and the dipole density $\hat{n}_{d,j} = \hat{q}_{d,j} + n_d$ (where $n_d \in \mathbb{N}$ is a suitable integer average dipole density, such that the local dipole density $n_{d,j}$ is non-negative) allows a dual description at low energies. In this section, we discuss the connection in terms of the continuum field theories that govern the ground state physics at integer filling. To understand this interplay, we first introduce a bosonic counting field $\phi(x)$ and a related phase field $\theta(x)$ following standard bosonization techniques~\cite{giamarchi:2003}. They fulfill canonical commutation relations of the form
\begin{equation}
    [\partial_x\phi(x), \theta(x^\prime)] = -i\pi \delta(x - x^\prime)
    \label{eq:CanCom}
\end{equation}
Continuum operators such as the particle density $n(x)$ and the single particle creation operator $b^\dagger(x)$ can be expressed in terms of these two fields:

\begin{equation}
    \begin{aligned}
        n(x) & = \bigl[n-\frac{1}{\pi}\partial_x \phi(x) \bigr]\sum_{m\in \mathbb{Z}}e^{2im\bigl(\pi n x-\phi(x)\bigr)}, \\ b^\dagger(x) & = \sqrt{n(x)} e^{-i \theta(x)},
    \end{aligned}
    \label{eq:Operators}
\end{equation}
where $n$ is the average particle density. In the continuum, the dipole-charge density relation becomes $\partial_x n_d(x) = n(x)$. The dipoles can also be expressed in a respective counting field $\phi_d(x)$ and a phase field $\theta_d(x)$. The counting field inherits the differential relation between the densities. A partial integration of the commutation relation (\ref{eq:CanCom}) then establishes the connection between the two pictures as

\begin{equation}
        \begin{split}
        \phi(x) &= \partial_x \phi_d(x) \\
        \partial_x \theta(x) &= -\theta_d(x) \,.
    \end{split}
    \label{eq:ParticleDipole}
\end{equation}
With this relation, one can express all low-energy theories either in particle or in dipole degrees of freedom. We note that the dipole density $n_d(x)$ and the dipole creation operator $d^\dagger(x)$ can be expressed in the same fashion as in \eq{eq:Operators} by replacing the charge fields with dipole fields.

The naive approach to obtain a low-energy field theory of the dipolar Bose-Hubbard model consists of performing a gradient expansion of the Hamiltonian (\ref{eq:DBH}), keeping only the lowest order terms~\cite{gorantla:2022,lake:2023}. The resultant continuum Hamiltonian is then, in each respective picture,

\begin{equation}
    \begin{aligned}
        H_{\text{Lif}}  &= \frac{v}{2 \pi} \int dx \left( K (\partial_x^2 \theta^2) + \frac{1}{K} (\partial_x \phi)^2 \right) \\ &= \frac{v}{2 \pi} \int dx \left( K (\partial_x \theta_d)^2 + \frac{1}{K} (\partial_x^2 \phi_d)^2 \right).
    \end{aligned}
    \label{eq:LifshitzAppendix}
\end{equation}
This is the one-dimensional version of the quantum Lifshitz theory, fully described by the two parameters $v$ and $K$ and gapless in both charges and dipoles. The higher derivatives in $\theta(x)$ compared to more common field theories follow from the quenched hopping term which enforces dipole symmetry, here manifest in the invariance under $\theta(x) \rightarrow \theta(x) + a + bx$. The $z = 2$ dynamical exponent of this theory furthermore leads to a dispersion of $\omega = v k^2$.

Lattice effects can destabilize such a theory at rational fillings. A renormalization group analysis shows that the operator $e^{i\phi(x)}$ has long-range correlation for all possible parameter values in the Lifshitz theory (\ref{eq:LifshitzAppendix}), implying that the cosine terms in the Hamiltonian which have to be added due to the lattice structure are always relevant and need not be neglected. At commensurate fillings $n=p/q$, one therefore has to consider an additional interaction term of the form
\begin{equation}
    g \cos{(2q\phi)}.
\end{equation}
This term gaps out the charges and spoils the emergence of the Lifshitz model at all rational fillings on general grounds.

\begin{figure*}
    \includegraphics[width=\textwidth]{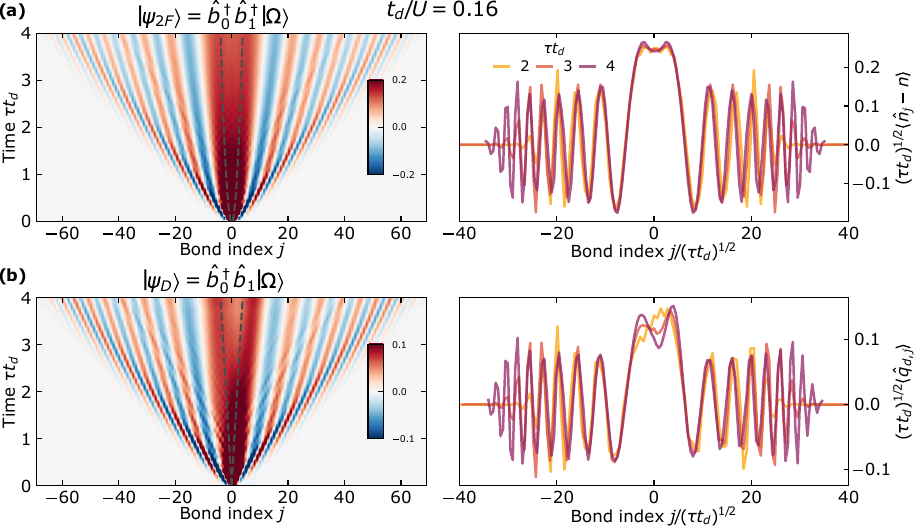}
    \caption{\label{fig:DipoleAppendix}
        \textbf{Lifshitz dynamics deep in the Luttinger liquid.} Time evolution of the two-fracton excitation $\ket{\psi_{2F}}$ in the density profile $\langle \hat{n}_j -n\rangle$ (a) and the dipole excitation $\ket{\psi_D}$ in the dipole charge profile $\langle \hat{q}_{d, j} \rangle$ (b) deep in the Luttinger liquid at $t/U = 0.16$. As the charge gap $\Delta_c$ is much smaller in this region, the Lifshitz theory (\ref{eq:LifshitzAppendix}) is a valid description on all numerically accessible time scales. Due to the duality of the Lifshitz theory in charge and dipole fields, both states exhibit the characteristic oscillatory diffusive modes. The Luttinger velocity $u_d$ is much slower than these modes, pushing the emergence of the low-energy ballistic dynamics to later times.
}
\end{figure*}

The cosine is the most relevant term in the full Hamiltonian. One can therefore safely expand it to obtain a Hamiltonian of the form (setting $q = 1$ for simplicity)
\begin{equation}
    H = \int dx \left\{ \frac{v}{2\pi} \left[ \frac{1}{K} (\partial_x \phi)^2 + K(\partial_x^2\theta)^2 \right] + 2g\phi^2\right\}.
    \label{eq:massiveLifshitz}
\end{equation}
This theory formally looks like a Lifshitz theory with an additional mass term for the charge density field $\phi(x)$. Yet, its spectrum is still gapless:

\begin{equation}
    \omega = v |k| \sqrt{k^2 + 4\pi g K/v}.
    \label{eq:mLifSpectrum}
\end{equation}
While the mass term changes the dynamical exponent to $z = 1$, it still allows gapless dipole excitations. In fact, using the relation $\phi^2(x) = (\partial_x \phi_d(x))^2$ and neglecting the higher-order term $(\partial_x \phi(x))^2 = (\partial^2_x \phi_d(x))^2$, one can see how this mass term induces the Hamiltonian of the dipolar Luttinger liquid
\begin{equation}
    H_{\text{LL}} = \frac{1}{2\pi} \int dx \left\{ u_d K_d \left(\partial_x\theta_d\right)^2 + \frac{u_d}{K_d} \left(\partial_x\phi_d\right)^2 \right\},
    \label{eq:LLAppendix}
\end{equation}
where $u_d = \sqrt{4\pi gKv}$ and $K_d=\sqrt{Kv/4\pi g}$. Further lattice cosines for the dipole field $\phi_d(x)$ can then be added, resulting in the sine-Gordon model (\ref{eq:SGH}). In this case, however, the cosine is only relevant for $K_d < 2$, thereby allowing the stabilization of the Luttinger liquid in the system.

While any non-zero $g$ therefore destabilizes the Lifshitz model in favor of the Luttinger liquid, at early-times Lifshitz-like physics may still emerge. While for low momenta $k \ll \sqrt{4\pi gK/v}$, the dispersion (\ref{eq:mLifSpectrum}) indeed results in the typical linear Luttinger relation $\omega = u_d |k|$, high momenta $k \gg \sqrt{4\pi gK/v}$ are well approximated by the Lifshitz prediction $\omega = v k^2$. As the charge gap $\Delta_c$ is proportional to the coupling $\Delta_c \sim \frac{g}{Kv}$, momenta above this energy scale effectively behave as if the system were gapless. Numerical studies confirm that that the charge gap $\Delta_c$ rapidly approaches small values at filling $n = 2$~\cite{zechmann:2023}. This implies that a large part of the spectrum follows the quadratic Lifshitz prediction~\cite{zechmann:2023a}. Correspondingly, the Lifshitz theory describes the time evolution of local excitations adequately up to very late times, at which point the low-momentum Luttinger modes finally enforce ballistic spreading.

\subsection{C. Dynamics of local dipole excitations}
The time evolution of a single localized excitation in the above quadratic field theories is directly accessible in analytical terms. For an initial local dipole excitation created by $\hat{d}^\dagger(x)$, we are interested in the dynamics of the dipole charge $q_d(x,\tau)=n_d(x,\tau)-n_d$.
Expanding the expressions in \eq{eq:Operators} to lowest order, we obtain

\begin{equation}
    \begin{aligned}
    q_d(x, \tau) = & \langle d(0, 0)(n_d(x, \tau) - n_d) d^\dagger(0,0) \rangle \\ \sim & -\frac{n_d}{\pi} \langle e^{i\theta_d(0, 0)} \partial_x \phi_d(x, \tau) e^{-i \theta_d(0,0)} \rangle.
    \end{aligned}
    \label{eq:DipoleChargeLowestOrder}
\end{equation}
As the theories in question are non-interacting, we can diagonalize them by going to an oscillator representation in momentum modes. The commutation relations (\ref{eq:CanCom}) imply that the Fourier modes $f_k = \frac{1}{\sqrt{L}} \int dx e^{-ikx} f(x)$ of the fields can be expressed in raising and lowering operators $a_k^{(\dagger)}$:

\begin{equation}
    \begin{aligned}
       (\partial_x \phi_d)_k & = \frac{1}{\sqrt{2}} A(k) (a_k + a_{-k}^{\dagger}) \\ \frac{1}{\pi} \theta_{d,k} & = \frac{i}{\sqrt{2}} \frac{1}{A(k)} (a_k - a_{-k}^{\dagger}),
    \end{aligned}
    \label{eq:OscillatorExpansion}
\end{equation}
where $a_k^{(\dagger)}$ follow the standard momentum mode relations and $A(k) = A(-k)$ is a non-universal pre-factor which ensures normalization. All quadratic Hamiltonians we consider are diagonal in the new operators, taking on the form

\begin{equation}
    H = \sum_k \omega(k) \left( a_k^{\dagger}a_k + \frac{1}{2} \right),
\end{equation}
where $\omega(k)$ is the relevant dispersion. In particular, the time evolution of the creation and annihilation operators is given by $a_k^{(\dagger)}(\tau) = e^{\pm i \omega(k) \tau} a_k^{(\dagger)}$.

The expectation value (\ref{eq:DipoleChargeLowestOrder}) can be successively simplified by going to momentum space and using the introduced representation. As operators at different momenta $k$ always commute, we can treat each $k$ mode separately by splitting up the exponential functions and commuting factors whose $k$ values differ from the density field momentum. The only non-trivial contributions that remain are

\begin{widetext}
\begin{equation}
    \begin{aligned}
    q_d(x, \tau) =  -\frac{n_d}{\pi\sqrt{2L}}  \sum_k A(k) \Bigl( e^{i(kx + \omega(k) \tau)} \langle e^{\frac{\pi}{\sqrt{2L}A(k)} a_k^{\dagger}} a_k e^{-\frac{\pi}{\sqrt{2L}A(k)} a_k^{\dagger}} \rangle + e^{i(kx - \omega(k) \tau)} \langle e^{-\frac{\pi}{\sqrt{2L}A(k)} a_k} a_k^\dagger e^{\frac{\pi}{\sqrt{2L}A(k)} a_k} \rangle \Bigr).
    \end{aligned}
    \label{eq:DipoleChargeSimplified}
\end{equation}
\end{widetext}
These terms can be evaluated by a Taylor expansion of the exponentials. Only zeroth- and first-order contributions are of relevance, as higher orders vanish due to different numbers of raising and lowering operators. The final expression is

\begin{equation}
    \begin{aligned}
    q_d(x, \tau) = & \frac{n_d}{L} \sum_k e^{ikx} \cos{(\omega(k) \tau)} \\ \; \rightarrow & n_d \int \frac{dk}{2\pi} e^{ikx} \cos{(\omega(k) \tau)},
    \end{aligned}
\end{equation}
where in the second line we go to a continuum limit. This can be seen as a superposition of plane waves with dispersion $\omega(k)$. For the Luttinger liquid (\ref{eq:LLAppendix}) with linear dispersion $\omega(k) = u_d |k|$, this naturally amounts to two ballistic counter-propagating modes

\begin{equation}
    q_d(x, \tau) = \frac{n_d}{2} \left[ \delta(x- u_d \tau) + \delta(x+ u_d \tau) \right]
\end{equation}
while for the quadratic Lifshitz mode the scaling function (\ref{eq:LifshitzScaling}) is reproduced. Dispersions which go faster than linear, such as in the Lifshitz model, lead to superluminal behavior in the continuum limit, as the velocity of high-momentum modes grows unbounded. In actual microscopic lattice Hamiltonians, a finite bandwidth $\sim 1/a$ where $a$ is the lattice spacing is present which prevents such unphysical behavior to arise. In continuum momentum integrals, this can be mimicked by a high-momentum cutoff of the order of the bandwidth.

For the massive Lifshitz model (\ref{eq:massiveLifshitz}) with the spectrum as given in \eq{eq:mLifSpectrum}, this derivation predicts the emergence the characteristic diffusive Lifshitz oscillations at early times, before the two Luttinger modes arise which push the oscillations in front of them and become the dominant feature at later times. The charge mass $\Delta_c$ is biggest close to the phase transition into the Mott insulator; in \figc{fig:Dipole}{b} the emergence of the Luttinger modes is already visible on accessible time scales. Deeper in the Luttinger liquid, the charge gap $\Delta_c$ rapidly decreases: we show the dynamics of the two-fracton and the dipole state at $t/U = 0.16$ in \fig{fig:DipoleAppendix}. As the Lifshitz theory is valid for much longer time scales, the respective density profiles in charges and dipole charges both exhibit the diffusive Lifshitz modes.

\subsection{D. Schrieffer-Wolff transformation in tilted Bose-Hubbard chains}
In this section, we restate the emergence of the dipolar Bose-Hubbard model as an effective early-time description of a tilted Bose-Hubbard chain. The derivation here follows a scheme already discussed in the literature~\cite{scherg:2021,moudgalya:2022, kohlert:2023, lake:2023}.

The dipole-moment conserving dynamics in the presence of a linear tilt can be made explicit by the application of a Schrieffer-Wolff (SW) transformation to the Hamiltonian (\ref{eq:HamiltonianTilt})~\cite{bravyi:2011}. Most generally, the SW transformation is applied to a Hamiltonian $\hat{H}$ which can be split into an already diagonal part $\hat{H}_0$ and an off-diagonal part $\hat{V}$ which serves as the perturbation:

\begin{equation}
    \hat{H} = \hat{H}_0 + \lambda \hat{V}.
    \label{eq:SWStart}
\end{equation}
Here, $\lambda$ is the coupling strength of the perturbation and is assumed to be small. The goal is to find a unitary transformation that diagonalizes the Hamiltonian $\hat{H}$ to some order $\mathcal{O}(\lambda^n)$. This transformation can be written as

\begin{equation}
    \hat{H}_{\text{eff}} = e^{\hat{S}} \hat{H} e^{-\hat{S}}
    \label{eq:SW}
\end{equation}
where $\hat{S}$ is an anti-hermitian operator. Especially in our case, it is paramount to note that the Hamiltonian $\hat{H}_{\text{eff}}$ is diagonal in a rotated basis which is connected to the original computational basis by

\begin{equation}
    \ket{\textbf{n}^{\prime}} = e^{\hat{S}} \ket{\textbf{n}}.
    \label{eq:SWBasis}
\end{equation}
One can expand the transformation (\ref{eq:SW}) using the Baker-Campbell-Hausdorff formula, which yields for the first few terms

\begin{equation}
    \hat{H}_{\text{eff}} = \hat{H} + [\hat{S}, \hat{H}] + \frac{1}{2} [\hat{S}, [\hat{S}, \hat{H}]] + \frac{1}{6} [\hat{S}, [\hat{S}, [\hat{S}, \hat{H}]]] + ...
    \label{eq:BCH}
\end{equation}

The final step to obtaining a controllable expression is to also expand the generator of the transformation $\hat{S}$ in orders of $\lambda$, where the zeroth-order term vanishes as there are no off-diagonal terms in $H$ at order $\mathcal{O}(\lambda^0)$:

\begin{equation}
    \hat{S} = \lambda \hat{S}_1 + \lambda^2 \hat{S}_2 + \lambda^3 \hat{S}_3 + \mathcal{O}(\lambda^4)
    \label{eq:SExpansion}
\end{equation}
This expansion is now plugged into \eq{eq:BCH} and all terms are organized in orders of $\lambda$. At each order, the $\mathcal{O}(\lambda^n)$ component of $\hat{S}$ is determined successively from the lower orders by enforcing that all off-diagonal terms vanish. The effective Hamiltonian is calculated from the remaining commutators at the respective order.

We apply the Schrieffer-Wolff transformation to a Bose-Hubbard chain in the limit of a strong tilt $\Delta \gg t,U$ to obtain an effective Hamiltonian to a certain order in $t/\Delta$ and $U/\Delta$. The goal is to eliminate the dipole moment violating hopping term. Prethermalization arguments suggest that the center of mass conserving dynamics describe the system adequately up to an exponentially long timescale $\tau_{\text{pre-th}} \sim \exp{\left( \Delta/t \right)}$, as we can always write a Schrieffer-Wolf Hamiltonian conserving dipole moment to an arbitrary order $\left( t/\Delta \right)^n$. This can be more physically interpreted as a severe restriction of the space of possible moves due to energy conservation: While dipole moment conserving processes leave the dominant tilt energy unchanged, any other process implies dissipation of tilt energy by means of the kinetic part of the Hamiltonian, which can only be achieved by a complicated multi-particle scattering~\cite{rosch2008, khemani:2022}.

We split the Hamiltonian of the tilted system into different parts:
\begin{equation}
    \begin{aligned}
    \hat{H} =  - &t \sum_j \left(\hat{b}^\dagger_j \hat{b}_{j+1} + \textrm{h.c.} \right) + t\left(\frac{U}{2t}\right) \sum_j \hat{n}_j (\hat{n}_j -1) \\ + & \Delta \sum_j j \hat{n}_j = t \hat{H}_{\text{kin}} + t \hat{H}_U + \hat{H}_\Delta
    \end{aligned}
    \label{eq:HamiltonianTiltAppendix}
\end{equation}
The eigenbasis of the diagonal Hamiltonian $\hat{H}_\Delta$ are the Fock states which have a well-defined particle number $N$ and dipole moment $P$.

Before stating the explicit form of the first orders of the generator (\ref{eq:SExpansion}), we simplify the expression for the effective Hamiltonian (\ref{eq:BCH}). The first order of the effective Hamiltonian, achieved by combining \eq{eq:BCH} with \eq{eq:SExpansion}, is

\begin{equation}
    \hat{H}_{\text{eff},1} = \hat{H}_\Delta + t \left(\hat{H}_{\text{kin}} + \hat{H}_U + [\hat{S}_1, \hat{H}_\Delta] \right)
    \label{eq:SWFOAppendix}
\end{equation}
The commutator $[\hat{S}_1, H_\Delta]$ has to be chosen such that it cancels all terms off-diagonal in the dipole moment. At first order, this is simply the kinetic term, which enforces the first-order condition for the SW generator
\begin{equation}
    [\hat{S}_1, \hat{H}_\Delta] = - \hat{H}_{\text{kin}}.
    \label{eq:SWFOCondAppendix}
\end{equation}
To this order, the Hamiltonian consists of the static terms governing the Hubbard interaction and the tilt. This is to be expected as no dipole-conserving process is possible at linear order in $t$; a single particle hopping always changes the dipole moment.

The second order of the expansion is:
\begin{equation}
    \begin{aligned}
    \hat{H}_{\text{eff}, 2} = & t^2 \left( \frac{1}{2} [\hat{S}_1, [\hat{S}_1, \hat{H}_\Delta] + [\hat{S}_1, \hat{H}_{\text{kin}} + \hat{H}_U] +[\hat{S}_2, \hat{H}_\Delta]\right) \\ = & t^2 \left( \frac{1}{2} [\hat{S}_1, \hat{H}_{\text{kin}}] + [\hat{S}_1,\hat{H}_U] +[\hat{S}_2, \hat{H}_\Delta]\right)
    \end{aligned}
    \label{eq:SWFOSecond}
\end{equation}
Here again, the commutator $[\hat{S}_2, \hat{H}_\Delta]$ is chosen in such a way as to cancel all off-diagonal contributions from the other terms. We can achieve this by investigating the structure of all appearing operators: As $\hat{H}_{\text{kin}}$ is completely off-diagonal and $\hat{H}_\Delta$ completely diagonal, the condition (\ref{eq:SWFOCondAppendix}) allows us to define a completely off-diagonal $\hat{S}_1$. This then implies that the commutator $[\hat{S}_1,\hat{H}_U]$ is off-diagonal as well and must be canceled; the second commutator $[\hat{S}_1, \hat{H}_{\text{kin}}]$ may contain both diagonal and off-diagonal contributions.
Defining a projector $\mathcal{P}$ that cancels all off-diagonal components, we thus arrive at a second-order condition of the form
\begin{equation}
    [\hat{S}_2, \hat{H}_\Delta] = -[\hat{S}_1,\hat{H}_U] - \frac{1}{2} (1 - \mathcal{P}) [\hat{S}_1, \hat{H}_{\text{kin}}] (1 - \mathcal{P})
    \label{eq:SWFOSecondCond}
\end{equation}
which fixes the second-order part of the effective Hamiltonian as 

\begin{equation}
    \hat{H}_{\text{eff}, 2} = \frac{t^2}{2} \mathcal{P} [\hat{S}_1, \hat{H}_{\text{kin}}] \mathcal{P}.
    \label{eq:SWSecondFinal}
\end{equation}
Finally, we  also  evaluate the third-order term, which is, after collecting all terms and applying the above conditions:

\begin{widetext}
\begin{equation}
    \begin{aligned}
    \hat{H}_{\text{eff}, 3} = & t^3 \left( \frac{1}{6} [\hat{S}_1, [\hat{S}_1, [\hat{S}_1, \hat{H}_{\Delta}]]] +\frac{1}{2} \left( [\hat{S}_2, [\hat{S}_1,\hat{H}_{\Delta}]] + [\hat{S}_1, [\hat{S}_2, \hat{H}_\Delta]] + [\hat{S}_1, [\hat{S}_1, \hat{H}_{\text{kin}} + \hat{H}_U]] \right) + [\hat{S}_2, \hat{H}_{\text{kin}} + \hat{H}_U] + [\hat{S}_3, \hat{H}_\Delta] \right) \\  
    = & t^3 \left( \frac{1}{3} [\hat{S}_1, [\hat{S}_1, \hat{H}_{\text{kin}}]] +\frac{1}{2} [\hat{S}_2, \hat{H}_{\text{kin}}] + [\hat{S}_2, \hat{H}_U] + [\hat{S}_3, \hat{H}_\Delta] \right).
    \end{aligned}
    \label{eq:SWThird}
\end{equation}
\end{widetext}

The structure of the commutator relation (\ref{eq:SWSecondFinal}) again makes a completely off-diagonal choice for $\hat{S}_2$ possible. Then, the commutator $[\hat{S}_2, \hat{H}_U]$ is also off-diagonal and needs to be canceled, while the first two commutators in the last line of $\hat{H}_{3, \text{eff}}$ can have both diagonal and off-diagonal components. The relevant condition reads
\begin{equation}
    \begin{aligned}
    [\hat{S}_3, \hat{H}_\Delta] = & - [\hat{S}_2, \hat{H}_U] - \frac{1}{2} (1- \mathcal{P})[\hat{S}_2, \hat{H}_{\text{kin}}] (1- \mathcal{P}) \\ & - \frac{1}{3} (1- \mathcal{P}) [\hat{S}_1, [\hat{S}_1, \hat{H}_{\text{kin}}]] (1- \mathcal{P}),
    \end{aligned}
    \label{eq:SWThirdCondition}
\end{equation}
which leads to a final expression for the third-order Hamiltonian as

\begin{equation}
    \hat{H}_{\text{eff}, 3} = t^3 \mathcal{P} \left( \frac{1}{2} [\hat{S}_2, \hat{H}_{\text{kin}}] + \frac{1}{3} [\hat{S}_1, [\hat{S}_1, \hat{H}_{\text{kin}}]] \right) \mathcal{P}
    \label{eq:SWThirdFinal}
\end{equation}
This order will suffice to obtain the dipole-conserving Bose-Hubbard model. We can therefore write down our effective Hamiltonian in cubic order:
\begin{widetext}
\begin{equation}
    \hat{H}_{\text{eff}} =  \hat{H}_\Delta + t \hat{H}_U + \frac{t^2}{2} \mathcal{P} [\hat{S}_1, \hat{H}_{\text{kin}}] \mathcal{P} + t^3 \mathcal{P} \left( \frac{1}{2} [\hat{S}_2, \hat{H}_{\text{kin}}] + \frac{1}{3} [\hat{S}_1, [\hat{S}_1, \hat{H}_{\text{kin}}]] \right) \mathcal{P} + \mathcal{O}(t^4)
    \label{eq:EffHamSWComplete}
\end{equation}
\end{widetext}
What remains to be specified is the precise structure of $\hat{S}_1$ and $\hat{S}_2$. The explicit calculation of these operators primarily consists of the evaluation of several commutation relations. Here, we limit ourselves to stating the results. The first order, determined by \eq{eq:SWFOCondAppendix}, can be expressed as

\begin{equation}
    \hat{S}_1 = \frac{1}{\Delta} \sum_j \hat{b}_j^\dagger \hat{b}_{j+1} - \hat{b}_{j+1}^\dagger \hat{b}_j
    \label{eq:S1}
\end{equation}
In particular, this also leads to a vanishing second order in the effective Hamiltonian, as $[\hat{S}_1, \hat{H}_{\text{kin}}] = 0$. In the non-interacting case $U = 0$, this is actually even more severe; as $[\hat{S}_1, [\hat{S}_1, \hat{H}_\Delta + \hat{H}_{\text{kin}}]] = 0$, the SW transformation stops at linear order, which allows a closed expression of the effective Hamiltonian and the rotated basis which diagonalizes it. The effective Hamiltonian then simply amounts to $\hat{H}_{\text{eff}} = \hat{H}_\Delta$.

This implies that the interactions are necessary to allow dipolar physics. For $U \neq 0$, the SW transformation does not stop abruptly and continues to all orders. The second order contribution $\hat{S}_2$ can be obtained from \eq{eq:SWFOSecondCond} by a lengthy calculation, which results in a form of
\begin{equation}
    \hat{S}_2 = - \frac{U}{2\Delta^2t} \sum_j \{\hat{n}_j, \hat{b}_{j-1}^\dagger \hat{b}_j - \hat{b}_j^\dagger \hat{b}_{j-1} - \hat{b}_j^\dagger \hat{b}_{j+1} - \hat{b}_{j+1}^\dagger \hat{b}_j\}
\end{equation}
where $\{\hat{A}, \hat{B} \} $ is the anti-commutator between two operators $\hat{A}$ and $\hat{B}$. Plugging this into the equation for the effective Hamiltonian (\ref{eq:EffHamSWComplete}) returns a dipole moment conserving Bose-Hubbard model with an additional nearest-neighbor interaction:
\begin{widetext}
    \begin{equation}
            \hat{H}_{\text{eff}} =  -\frac{t^2 U}{\Delta^2} \sum_j \left( \hat{b}_j^{\dagger} \hat{b}_{j+1}^2 \hat{b}_{j+2}^{\dagger} + \textrm{h.c.} \right) + \left(\frac{U}{2} - \frac{2t^2U}{\Delta^2}\right) \sum_j \hat{n}_j (\hat{n}_j -1) + \frac{2t^2 U}{\Delta^2} \sum_j \hat{n}_j \hat{n}_{j+1} + \sum_j \left(j\Delta - \frac{2t^2U}{\Delta^2} \right) \hat{n}_j.
    \label{eq:SWTiltCubicAppendix}
    \end{equation}
\end{widetext}
This shows that dipolar physics is indeed realized in the prethermal dynamics of an interacting tilted lattice.

\begin{figure*}
    \includegraphics[width=\textwidth]{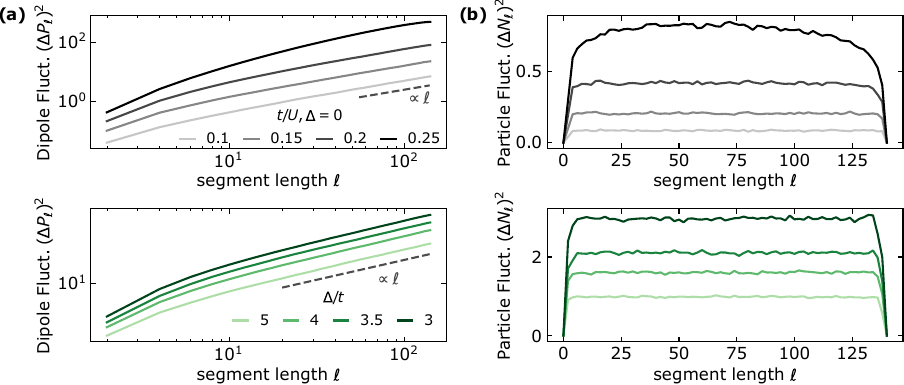}
    \caption{\label{fig:Static}
        \textbf{Static dipole and particle fluctuations.} We show the dipole (a) and particle (b) fluctuations both for a Mott insulator at unity filling $n = 1$ in a regular Bose-Hubbard chain (gray lines) and for a state that is adiabatically prepared over 20 hopping times as described in the main text (green lines). In both cases, the dipole fluctuations exhibit a linear dependence on the segment size, impeding a simple distinction on the basis of static fluctuations, as the finite extent of the Wannier-Stark orbitals in the tilted lattice modifies the expected scaling. Similarly, the charge fluctuations are constant in all cases as charges remain gapped both in the regular Mott insulator as well as in the effective Hamiltonian of the tilted lattice.
    }
\end{figure*}

However, one has to be careful when comparing this expectation to the actual properties of the system. As stated in the beginning of the section, the Hamiltonian realizes the desired form not in the original basis, but rather in a dressed basis obtained by the rotation (\ref{eq:SWBasis}). This renders certain quantities much more opaque, as we sketch in the following.

\subsection{E. Dipole moment fluctuations in tilted systems}
In the non-interacting case $U = 0$ a closed expression can be found, where the effective Hamiltonian keeps the form of the tilt contribution, $\hat{H}_{\text{eff}} = \hat{H}_\Delta$. This is due to the fact that the SW transformation stops at linear order, $\hat{S} = \hat{S}_1$. It is immediately clear that due to the lack of interactions, this can be treated as a single-particle problem in a tilted chain. This is known as Wannier-Stark localization, where the presence of a tilt of arbitrary strength $\Delta \neq 0$ leads to localization of the eigenstates to a lattice site, a drastic change compared to the plane wave eigenfunctions in the $\Delta = 0$ case~\cite{wannier1960}. A sharply localized orbital centered around a lattice site $i$ is mapped to the respective Wannier-Stark orbital, which, while still strongly localized, exhibits a finite extension. Far away from the central site, the asymptotic behavior features a super-exponential decay
\begin{equation}
   |\psi_j| \sim \exp\left( - |j-i| \log \left( \frac{|j-i|\Delta}{e t}\right) \right),
   \label{eq:Wannier}
\end{equation}
where $\psi_j$ is the wave function amplitude at site $j$ and $e$ is the Euler number. We see that a characteristic length scale $t/\Delta$ is introduced over which the particle is delocalized. To first order, this is also the mapping in the fully interacting case; the effective dipolar Hamiltonian \eq{eq:SWTiltCubicAppendix} is therefore expressed in the Wannier-Stark orbitals, instead of the exactly localized lattice site states.

In the main text, we have argued that this prevents a distinction of the dipolar ground states from a generic Mott insulator by means of static measurements of dipole moment fluctuations $\bigl(\Delta P_\ell\bigr)^2$ in subsegments of size $\ell$ of a tilted chain. These are defined as

\begin{equation}
    \bigl(\Delta P_\ell\bigr)^2 = \left\langle \left(\hat{P}_\ell - \langle \hat{P}_\ell \rangle\right)^2 \right\rangle,
    \label{eq:DipoleFluctAppendix}
\end{equation}
where $\hat{P}_\ell=\sum_{j=1}^\ell \hat{q}_{d,j-\ell/2}$ measures the dipole moment in the segment of size $\ell$. However, even if we achieve a state that is close to the desired ground state by an adiabatic preparation (as our derivation of the effective Hamiltonian (\ref{eq:SWTiltCubicAppendix}) suggests, provided it is slow enough), the relevant state conserves the dipole moment not in the real-space measurement basis, but in the space of Wannier-Stark Fock states. In these, the dipole fluctuations in a segment of size $\ell$ follow the expected scaling laws: They are constant in the dipole Mott insulator as the dipole degrees are gapped, and increase logarithmically in segment size in the gapless Luttinger liquid. However, the Wannier-Stark orbitals are smeared out over a finite length $\propto t/\Delta$, which implies further corrections to the scaling. Concretely, as the density is constant provided we are far away from the edges of the system, the number of particles in an average Fock state picked from a snapshot measurement is proportional to the segment length $\ell$. As each of those contributes constant dipole moment fluctuations due to the extension (\ref{eq:Wannier}), the dipole fluctuations in the measurement basis actually scale as $\bigl(\Delta P_\ell\bigr)^2 \sim \ell$. This is the same scaling as one would expect from a regular Mott insulator: In such a state, we would expect a finite and constant density of particle-hole excitations due to the finite charge gap, each of which carries a dipole charge of $\pm 1$. The dipole moment in the segment of a randomly picked Fock state should therefore follow a binomial distribution. Therefore, the dipole moment fluctuations should also increase proportionally to the segment size, $\bigl(\Delta P_\ell^{\text{MI}}\bigr)^2 \sim l$. In particular, we see that the effective realization introduces a subtlety that impedes a simple confirmation of the dipole character. We have confirmed this expectation using MPS simulations. In \figc{fig:Static}{a}, we show the dipole fluctuations as a function of segment size both for a regular Mott insulator at filling $n = 1$ in a standard Bose-Hubbard model with $\Delta = 0$ (gray lines), and for a state at filling $n = 2$ which has been prepared adiabatically in the tilted system as described in the main text (green lines). Both exhibit the same scaling, as we would expect from our discussion.

One can also look at the particle fluctuations in a segment
\begin{equation}
    \bigl(\Delta N_\ell\bigr)^2 = \left\langle \left(\hat{N}_\ell - \langle \hat{N}_\ell \rangle\right)^2 \right\rangle,
    \label{eq:ChargeFluctAppendix}
\end{equation}
where $\hat{N}_\ell=\sum_{j=1}^\ell \hat{n}_{j-\ell/2}$ is the particle number in the middle segment of size $\ell$. For this quantity, the Wannier-Stark localization should not imply a modification of the scaling law: Only at the edges of the segment might a particle escape or intrude due to its orbit's extension, which implies at most constant fluctuations. As the charge degrees of freedom are gapped both in the dipole-conserving model and in the regular Mott state, all should in general exhibit constant charge fluctuations $\bigl(\Delta N_\ell\bigr)^2 \sim \textrm{const}$. Indeed, an evaluation from the same snapshots as for the dipole fluctuations shows a constant value in both cases, \figc{fig:Static}{b}. While this does not confirm the presence of dipole-conserving physics, this result does confirm a finite gap for charges in the tilted system at early times, ruling out a regular Luttinger liquid of bosons as the realized phase in spite of the low interaction-to-hopping ratio. 

By contrast, dynamical probes as described in the main text do allow for a distinction between the regular Mott state and actual dipole ground states as the local processes that drive changes in the dipole moment are much less affected by the basis transformation.

\subsection{F. Details on numerical methods}
All our numerical data is obtained using Matrix Product States (MPS) as implemented in the TeNPy library~\cite{hauschild:2018}. For the simulation of the time evolution of local excitations in the explicitly dipole-moment conserving model \eq{eq:DBH}, we first compute the ground state in an infinite-size system at filling $n = 2$ for different hopping-to-interaction ratios $t/U$ using Density Matrix Renormalization Group (DMRG)~\cite{white:1992, white:1993, vidal:2007}. Besides the standard implementation of the $U(1)$ particle number conservation~\cite{singh:2010, singh:2011}, we also directly enforce dipole moment conservation to gain an additional computational speed-up~\cite{zechmann:2023} and use the subspace expansion method to avoid local minima~\cite{hubig:2015}. The unit cell of the state that is to be optimized is $L = 10$, which afterwards is enlarged to $L = 140$ by concatenating copies. Then, the relevant creation and annihilation operators are applied to obtain the sought-after low-energy excitation. We time-evolve these using the $W^{II}$ algorithm~\cite{zaletel:2015}, which can treat longer-range terms as present in this model. We fix a maximal local boson occupation of $n_{\text{max}} = 8$ to ensure converged results. The maximal bond dimension for both the ground state search and the time evolution is $\chi_\text{max} = 1600$. The results appear to be well-converged even in the gapless Luttinger liquid for the times considered.

The simulation of the proposed experimental scheme starts from a homogeneous product state of filling $n = 2$ in a finite system of size $L = 140$. The ramping process is modeled by a time evolution using the Time-Evolving Block Decimation (TEBD) algorithm~\cite{vidal:2004}, governed by the Hamiltonian \eq{eq:HamiltonianTilt} with a time-dependent hopping parameter $t(\tau)$ and fixed interaction strength $U$ and tilt strength $\Delta$. We start from the static case $t = 0$. As a product state is an eigenstate of the Hamiltonian with vanishing hopping parameter for arbitrary values of the tilt $\Delta$, we do not include the increase of the tilt in our simulation, as this would amount only to a complex phase. We increase $t$ in discrete steps after a certain number of updates by re-initializing the TEBD representation of the time evolution operator, until we reach the final value $t$ with $t/U = 2$ and the desired ratio $\Delta/t$. The ramping process takes place over twenty hopping periods in terms of the final hopping parameter, $t\tau = 20$. After applying a particle creation operator on the two adjacent sites in the middle of the chain, we further evolve the state with the now constant time evolution operator with hopping strength $t$. At fixed time steps, including just after the preparation of the excitation, we sample $N= 10\,000$ Fock states from the state by performing projective measurements, thereby obtaining a distribution for the local occupation numbers $n_j$. From this, we can obtain both the particle number fluctuations $\Delta N$ and the dipole moment fluctuations $\Delta P$ at these times for different segment sizes $l$. The size $l = 80$ for Fig.\ 3 in the main text is chosen such that we can be sure to consider only bulk effects, as the time evolution from a non-eigenstate leads to excitations emerging from the edges of the system. The maximal on-site particle number is $n_{\text{max}} = 6$, while the maximal bond dimension is $\chi_{\text{max}} = 600$. We also perform the same measurement scheme in the case that no particles are added before the fixed-$t$ time evolution. The dipole fluctuations in time in this case are subtracted from those of the excitation state. Thereby, we expect to consider only fluctuations arising from the dynamics, as further fluctuations that arise due to the imperfectness of the preparation scheme are canceled.

For comparison, we also calculate the ground state of the regular Bose-Hubbard model at filling $n = 1$ for different parameter sets in the Mott phase using DMRG, add two particles in the middle, and perform the same measurement scheme in a comparable time evolution. The results, with the pure ground state fluctuations subtracted, are shown in Fig.\ 3 of the main text as well.
\end{document}